\documentclass[onecolumn,10pt,a4paper,sort&compress,aps,pra,showpacs,superscriptaddress]{revtex4-2}

\usepackage[english]{babel}

\usepackage[caption=false]{subfig}
\addto\captionsenglish{\renewcommand{\figurename}{FIG.}}

\usepackage{overpic}
\usepackage{amsmath,amssymb,amsthm}
\usepackage{commath}
\usepackage{graphicx}
\usepackage{dcolumn}
\usepackage{bm}
\usepackage{hyperref}
\usepackage{etoolbox} 
\usepackage{siunitx}
\usepackage{multirow}
\DeclareSIUnit\Molar{\textsc{M}}
\usepackage[capitalise,nameinlink]{cleveref}
\newcommand{\aref}[1]{\hyperref[#1]{Appendix}}
\usepackage{blkarray}

\usepackage[normalem]{ulem}
\usepackage{comment}
\usepackage[usenames,dvipsnames]{color}

\renewcommand{\vec}[1]{\bm{#1}}
\newcommand{\intd}{\mathop{}\!\mathrm{d}}
\newcommand{\ds}{\Delta s}
\newcommand{\cross}{\times}
\newcommand{\pdiff}[2]{\frac{\partial{#1}}{\partial{#2}}}
\newcommand{\transpose}{\top}

\newcommand{\ct}{c_{\theta}}
\renewcommand{\cp}{c_{\phi}}
\newcommand{\cpsi}{c_{\psi}}
\newcommand{\st}{s_{\theta}}
\renewcommand{\sp}{s_{\phi}}
\newcommand{\spsi}{s_{\psi}}

\begin{document}


\title{Efficient simulation of filament elastohydrodynamics in three dimensions}

\author{Benjamin J. Walker}
\email{Corresponding author: benjamin.walker@maths.ox.ac.uk}
\affiliation{Wolfson Centre for Mathematical Biology, Mathematical Institute, University of Oxford, Oxford, OX2 6GG, UK}

\author{Kenta Ishimoto}
\email{ishimoto@kurims.kyoto-u.ac.jp}
\affiliation{Research Institute for Mathematical Sciences, Kyoto University, Kyoto, 606-8502, Japan}

\author{Eamonn A. Gaffney}
\email{gaffney@maths.ox.ac.uk}
\affiliation{Wolfson Centre for Mathematical Biology, Mathematical Institute, University of Oxford, Oxford, OX2 6GG, UK}


\date{\today}

\begin{abstract}
Fluid-structure simulations of slender inextensible filaments in a
viscous fluid are often plagued by numerical stiffness. Recent coarse-graining
studies have reduced the computational requirements of simulating such
systems, though have thus far been limited to the motion of planar filaments.
In this work we extend such frameworks to filament motion in three dimensions,
identifying and circumventing coordinate-system singularities introduced by
filament parameterisation via repeated changes of basis. The resulting
methodology enables efficient and rapid study of the motion of flexible
filaments in three dimensions, and is readily extensible to a wide range of
problems, including filament motion in confined geometries, large-scale active
matter simulations, and the motility of mammalian spermatozoa.

\end{abstract}

\maketitle

\section{Introduction}

The coupled elasticity and hydrodynamics of flexible inextensible filaments on
the microscale are of significance to much of biology, biophysics and soft
matter physics. For example, many organisms possess slender flagella or cilia,
utilised for driving flows and even locomotion, whilst investigation into the
role of synthetic filaments as both soft deformable sensors and methods of
propulsion has been the subject of recent enquiry
\cite{Gray1928,Hancock1953,Berg1973,Tornberg2004,Roper2006,Pozrikidis2010,Pozrikidis2011,Tottori2012,Meng2018}.
As a result, the complex mechanics of fluid-structure interaction has been
well-studied, utilising methods such as the slender body and resistive force
theories of \citet{Hancock1953,Gray1955,Johnson1977}, through to the exact
representations of boundary integral methods as used by
\citeauthor{Pozrikidis2010}
\cite{pozrikidis1992,Pozrikidis2010,Pozrikidis2011}. A fundamental barrier to
much numerical investigation has been the severe stiffness associated with the
equations of filament elasticity when coupled to viscous fluid dynamics.
Hence, as remarked in the recent and extensive review of \citet{DuRoure2019},
an appropriate framework, capable of realising efficient simulation of
filament elastohydrodynamics, is crucial for the numerical study of filament
mechanics.

Recently, significant progress has been made in resolving the dynamics of
planar filaments, with the work of \citet{Moreau2018} presenting a
coarse-grained model of filament elasticity that overcame much of the
stiffness previously associated with slender elastohydrodynamics. Key to this
approach was the integration of the pointwise force and moment balance
equations, the spatial discretisation of which yielded a relatively simple
system of ordinary differential equations to solve in order to describe
filament motion. Additionally demonstrated to be flexible in the original
publication of \citeauthor{Moreau2018}, this framework has been extended to
include non-local hydrodynamics in both infinite and semi-infinite domains,
and applied to a variety of single and multi-filament problems
\cite{Hall-McNair2019,Walker2019d}. However, being confined to two dimensions
limits the potential scope and applicability of these approaches, with three
dimensional filament motion being readily and frequently observed in a
plethora of biophysical systems, such as the complex flagellar beating found
in spermatozoa or the helically-driven monotrichous bacterium
\textit{Escherichia coli} \cite{Yanagimachi1970,Berg1973}.

However, for non-planar filaments in three dimensions there is currently no
methodology analogous to that of \citeauthor{Moreau2018}, with
state-of-the-art frameworks still plagued by extensive numerical stiffness,
necessitating costly computation to the extent that practical simulation
studies have been limited and parameter space studies are largely prohibited.
With three dimensions inherently more challenging than lower dimensional
settings, this field has seen developments such as the recent work of
\citet{Schoeller2019}, which utilises a quarternion representation of filament
orientation to parameterise the three dimensional shape of the slender body.
However, in this framework, numerical care is required to satisfy the
inextensibility condition, with similar such consideration necessary in the
earlier methodologies of
\citet{Olson2013,Simons2015,Ishimoto2018a,Bouzarth2011}, each of which are
equipped with non-local slender-body hydrodynamics and consider
nearly inextensible filaments. Consequently, these existing approaches often
require the use of sophisticated computing hardware in order to simulate
filament motion, with typical simulations of \citeauthor{Ishimoto2018a} having
a runtime of multiple hours on high performance computing clusters. The recent
work of \citet{Jabbarzadeh2020} compared and contrasted these nearly
inextensible approaches with a truly inextensible scheme, concluding that both
accuracy and efficiency was afforded by the latter in a range of biological
and biophysical modelling scenarios. Despite their improved efficiency,
typical walltimes for these filament simulations are measured on a timescale
of hours on typical hardware. Thus, there remains significant scope for the
development of an efficient framework for the simulation of
inextensible elastic filaments in three dimensions, one in which
filament dynamics can be rapidly computed on non-specialised hardware
on timescales of seconds or minutes, thus facilitating a wealth of
future studies and explorations into complex and previously intractable
biological and physical systems.

Hence, the fundamental objective of this study is to develop and describe an
efficient framework for the numerical simulation of filament mechanics in
three dimensions. We will build upon the recent and significant work of
\citet{Moreau2018}, extending their approach to include an additional spatial
dimension via a generalisation of the Frenet triad and integration of the
governing equations of elasticity. We will overcome fundamental issues with
simple single parameterisations of a filament in three dimensions, presenting
an effective computational approach utilising adaptive reparameterisation and
basis selection. We will then validate the presented framework by
consideration of three candidate test problems, simulating well-documented
behaviours of filaments in a viscous fluid and including a side-by-side
comparison against the existing and recent methodology of
\citet{Ishimoto2018a}. Finally, we will showcase the flexibility and general
applicability of the presented approach by describing and exemplifying a
number of methodological extensions.

\section{Methods}\label{sec:methods}
\subsection{Equations of elasticity}
We consider a slender, inextensible, unshearable filament in a viscous
Newtonian fluid, with its centreline described by $\vec{x}(s)$, parameterised
by arclength $s\in[0,L]$ for dimensional filament length $L$. We model the
filament as a Kirchhoff rod with arclength-independent material parameters,
circular cross-sections, and, in the first instance, no intrinsic curvature or
intrinsic torsion. Both the filament and fluid inertia are negligible for the
physical scales associated with many applications, especially those associated
with cellular flagella and cilia; hence there is no inertia here and
throughout. Along the filament we have the pointwise conditions of force and
moment balance, given explicitly by
\begin{align}
 	\vec{n}_s - \vec{f} &= \vec{0}\,, \label{eq:force_balance}\\
 	\vec{m}_s + \vec{x}_s\cross\vec{n} - \vec{\tau} &= \vec{0}\,, \label{eq:pointwise_moment_balance}
\end{align} 
for contact force and torque denoted $\vec{n},\vec{m}$ respectively and where
a subscript of $s$ denotes differentiation with respect to arclength. The
quantity $\vec{f}$ is the force per unit length applied on the fluid medium by
the filament, which we will later express in terms of the filament velocity
$\dot{\vec{x}}$, where here a dot denotes a time derivative. Similarly,
$\vec{\tau}$ is the torque per unit length applied on the fluid medium by the
filament. Whilst any standard boundary condition may be considered in the
formalism below, throughout we impose zero force and torque at the filament
ends so that
\begin{equation}
	\vec{n}(0)=\vec{n}(L)=\vec{m}(0)=\vec{m}(L)=\vec{0}\,.
\end{equation}
In the Kirchhoff framework note that the contact force is not constitutive, but
simply an undetermined Lagrange multiplier for the intrinsic constraints of
inextensibility and no shearing of the filament cross section in any direction
\cite{Antman2005}. Thus we may eliminate $\vec{n}(s)$ at the earliest
opportunity; using the boundary condition $\vec{n}(L)=0$ and
\cref{eq:force_balance} we have
\begin{equation}\label{eq:n_in_terms_of_f}
	\vec{n}(s)=-\int\limits_s^L \vec{f}(\tilde{s})\intd{\tilde{s}}\,.
\end{equation}
This relation is also subject to the constraint that the boundary condition
$\vec{n}(0)=\vec{0}$ is satisfied, generating a global force balance
constraint for the applied force per unit length,
\begin{equation}\label{eq:total_force_balance}
	\vec{0}=\int\limits_0^L \vec{f}(\tilde{s})\intd{\tilde{s}},
\end{equation}
which we carry forward into the formalism below together with the elimination
of $\vec{n}(s)$ via \cref{eq:n_in_terms_of_f}. Thus, and as originally
considered in \citet{Moreau2018}, integration of
\cref{eq:pointwise_moment_balance} and use of the boundary condition
$\vec{m}(L)=\vec{0}$, reveals the integrated moment balance
\begin{equation}\label{eq:total_moment_balance}
	- \int\limits_{s}^{L}\left[ (\vec{x}(\tilde{s})-\vec{x}(s))\cross\vec{f}(\tilde{s}) + \vec{\tau}(\tilde{s})\right]\intd{\tilde{s}} = \vec{m}(s)\,.
\end{equation}
Given a right-handed orthonormal director basis
$\{\vec{d}_1(s),\vec{d}_2(s),\vec{d}_3(s)\}$, generalising the Frenet triad
such that $\vec{d}_3$ corresponds to the local filament tangent, following
\citet{Nizette1999} we define the twist vector $\vec{\kappa}$ by
\begin{equation}
	\pdiff{\vec{d}_\alpha}{s} = \vec{\kappa}\cross\vec{d}_\alpha
\end{equation} 
for $\alpha=1,2,3$. Writing $\vec{\kappa}=\sum_{\alpha}\kappa_\alpha
\vec{d}_\alpha$, for bending stiffness $EI$ we use the Euler-Bernouilli
constitutive relation of the Kirchhoff formalism to relate the contact torque
$\vec{m}$ and the twist vector $\vec{\kappa}$ \cite{Antman2005}, via
\begin{equation}\label{eq:constitutive_moments}
	\vec{m} = EI\left(\kappa_1\vec{d}_1 + \kappa_2\vec{d}_2 + \frac{1}{1+\sigma}\kappa_3\vec{d}_3\right)\,,
\end{equation}
where $\sigma$ is the Poisson ratio \cite{Nizette1999}, assumed to be
constant. With this constitutive relation the integrated moment balance
equations in the $\vec{d}_\alpha$ directions are simply
\begin{equation}\label{eq:moment_balance}
	 -\vec{d}_\alpha(s)\cdot\int\limits_s^L \left[(\vec{x}(\tilde{s})-\vec{x}(s))\cross\vec{f}(\tilde{s}) + \vec{\tau}(\tilde{s})\right]\intd{\tilde{s}} = \frac{EI}{1+\delta_{\alpha,3}\sigma}\kappa_{\alpha}(s)\,,
\end{equation}
for $\alpha=1,2,3$ and where $\delta_{a,b}$ denotes the Kronecker delta.

\subsection{Filament discretisation}
In discretising the filament we follow the approach of \citet{Walker2019d}, as
previously applied to planar filaments and itself building upon the earlier
work of \citet{Moreau2018}. We approximate the filament shape with $N$
piecewise-linear segments, each of constant length $\ds$, with segment
endpoints having positions denoted by $\vec{x}_1,\ldots,\vec{x}_{N+1}$ and the
constraints of inextensibility and the absence of cross section shear are
satisfied inherently. The endpoints of the $i$\textsuperscript{th} segment
correspond to $\vec{x}_i$ and $\vec{x}_{i+1}$ for $i=1,\ldots,N$, with the
local tangent $\vec{d}_3$ being constant on each segment and denoted
$\vec{d}_3^{i}$. In what follows we will consider a discretisation of
$\vec{d}_1,\vec{d}_2$ such that they are also constant on each segment, and we
denote these constants similarly as $\vec{d}_1^i,\vec{d}_2^i$. Writing $s_i$
for the constant arclength associated with each material point $\vec{x}_i$, we
apply \cref{eq:moment_balance} at each of the $s_i$ for $i=1,\ldots,N$,
splitting the integral at the segment endpoints to give
\begin{equation}\label{eq:discretised_moment}
	- \vec{d}_\alpha^i\cdot\sum\limits_{j=i}^{N}\int\limits_{s_j}^{s_{j+1}}
	  \left[(\vec{x}(\tilde{s})-\vec{x}_i)\cross\vec{f}(\tilde{s})+\vec{\tau}(\tilde{s})\right]\intd{\tilde{s}}
	  = \frac{EI}{1+\delta_{\alpha,3}\sigma}\kappa_{\alpha}(s_i)\,,
\end{equation}
for $\alpha=1,2,3$. On the $j$\textsuperscript{th} segment, $\vec{x}$ may be
written as $\vec{x}(s) = \vec{x}_j + \eta(\vec{x}_{j+1} - \vec{x}_j)$, where
$\eta\in[0,1]$ is given by $\eta = (s-s_j)/\ds$. Additionally discretising the
force per unit length as a continuous piecewise-linear function, with $\eta$
as above we have $\vec{f}(s)=\vec{f}_j + \eta (\vec{f}_{j+1} -
\vec{f}_j)$ on the segment, where we write $\vec{f}_j=\vec{f}(s_j)$.
Substitution of these parameterisations into \cref{eq:discretised_moment} and
subsequent integration yields, after simplification,
\begin{equation}\label{eq:final_moment_balance}
	-\vec{d}_{\alpha}^i \cdot\left(\vec{I}_i^f + \vec{I}_i^{\tau}\right) = \frac{EI}{1+\delta_{\alpha,3}\sigma}\kappa_{\alpha}(s_i)\,,
\end{equation}
where the integral contribution of the force and torque densities are denoted $\vec{I}_i^f$ and $\vec{I}_i^\tau$ respectively. With this discretisation $\vec{I}_i^f$ has reduced to
\begin{equation}\label{eq:expanded_moment}
\vec{I}_i^f = \sum\limits_{j=i}^N \left\{\left[\frac{\ds}{2}\left(\vec{x}_j-\vec{x}_i\right) + \frac{\ds^2}{6}\vec{d}_3^j\right] \cross \vec{f}_j 
+\left[\frac{\ds}{2}\left(\vec{x}_j-\vec{x}_i\right) + \frac{\ds^2}{3}\vec{d}_3^j\right] \cross \vec{f}_{j+1}\right\} \,,
\end{equation}
in agreement with expressions for planar filaments found in
\citet{Moreau2018,Walker2019d}. As we will highlight below, the contributions
of the applied torque per unit length are relatively small given the
slenderness of the filament, motivating a less refined discretisation for
$\vec{I}_i^{\tau}$. Hence, taking the piecewise constant discretisation
$\vec{\tau}=\vec{\tau}_j$ on the $j$\textsuperscript{th} segment, we have the
simple expression
\begin{equation}
	\vec{I}_i^{\tau} = \sum\limits_{j=i}^N \ds\vec{\tau}_j\,.
\end{equation}
From the above we see explicitly that the integral component of each moment
balance equation may be written as a linear operator acting on the $\vec{f}_j$
and the $\vec{\tau}_j$. Similarly, with this piecewise-linear force
discretisation the integrated force balance of \cref{eq:force_balance} simply
reads
\begin{equation}\label{eq:final_force_balance}
	-\frac{\ds}{2}\sum\limits_{j=1}^{N}(\vec{f}_j+\vec{f}_{j+1}) = \vec{n}(0)\,.
\end{equation}
We write
$\vec{F}=[f_{1,x},f_{1,y},f_{1,z},\ldots,f_{N+1,x},f_{N+1,y},f_{N+1,z}]^{\transpose}$
for components $f_{j,x},f_{j,y,}f_{j,z}$ of $\vec{f}_j$ with respect to some
fixed laboratory frame with basis $\{\vec{e}_x,\vec{e}_y,\vec{e}_z\}$, and
similarly $\vec{T}$ for the vector of components of applied torque per unit
length. Here and throughout, forces and torques are written with respect to
the laboratory reference frame. With this notation, we may write the equations
of force and moment balance as
\begin{equation}\label{eq:BFR}
	-\mathcal{B}\begin{bmatrix}\vec{F}\\\vec{T}\end{bmatrix} = \vec{R}\,,
\end{equation}
where $\mathcal{B}$ is a matrix of dimension $(3N+3)\times(6N+3)$ with rows
$\mathcal{B}_k$. For $k=1,2,3$ the first $3N+3$ columns are given by
\begin{equation}
\begin{aligned}
	\mathcal{B}_1 & = \frac{\ds}{2}[1,0,0,2,0,0,2\ldots,2,0,0,1,0,0]\,,\\
	\mathcal{B}_2 & = \frac{\ds}{2}[0,1,0,0,2,0,0,2\ldots,2,0,0,1,0]\,,\\
	\mathcal{B}_3 & = \frac{\ds}{2}[0,0,1,0,0,2,0,0,2\ldots,2,0,0,1]\,,\\
\end{aligned}
\end{equation}
and correspond to the force balance of \cref{eq:final_force_balance}, with the
remaining $3N$ columns zero. The remaining rows of $\mathcal{B}$ encode the
moment balance of \cref{eq:final_moment_balance} as expanded in
\cref{eq:expanded_moment}, organised in triples such that
$\mathcal{B}_{3(i-1)+3+\alpha}$ projects the $i$\textsuperscript{th} moment
balance equation onto $\vec{d}_{\alpha}^i$, in that this
$(3i+\alpha)$\textsuperscript{th} row of $\mathcal{B}$ captures the
$\vec{d}_{\alpha}^i$ component of $-(\vec{I}_i^{f} +
\vec{I}_i^{\tau})$. The cross products inherited from
\cref{eq:expanded_moment} may now be notationally simplified by use of the
cyclic property of the scalar triple product, explicitly giving
\begin{align}
    \vec{d}_{\alpha}^i\cdot\vec{I}_i^{f} &= \sum\limits_{j=i}^N \left\{\vec{d}_{\alpha}^i\cdot\left[\frac{\ds}{2}\left(\vec{x}_j-\vec{x}_i\right) + \frac{\ds^2}{6}\vec{d}_3^j\right] \cross \vec{f}_j + \vec{d}_{\alpha}^i\cdot\left[\frac{\ds}{2}\left(\vec{x}_j-\vec{x}_i\right) + \frac{\ds^2}{3}\vec{d}_3^j\right] \cross \vec{f}_{j+1}\right\} \\
    &= \sum\limits_{j=i}^N \left\{\vec{d}_{\alpha}^i\cross\left[\frac{\ds}{2}\left(\vec{x}_j-\vec{x}_i\right) + \frac{\ds^2}{6}\vec{d}_3^j\right] \cdot \vec{f}_j + \vec{d}_{\alpha}^i\cross\left[\frac{\ds}{2}\left(\vec{x}_j-\vec{x}_i\right) + \frac{\ds^2}{3}\vec{d}_3^j\right] \cdot \vec{f}_{j+1}\right\}\,,
\end{align}
with the latter expression readily transcribed as a linear operator acting on
the $\vec{f}_j$ for $j=i,\ldots,N$. Analogously, we have 
\begin{equation}
	\vec{d}_{\alpha}^i\cdot\vec{I}_i^{\tau} = \ds\vec{d}_{\alpha}^i\cdot\sum\limits_{j=i}^N \vec{\tau}_j\,,
\end{equation}
from which a linear operator acting on the $\vec{\tau}_j$ for $j=i,\ldots,N$
can be constructed. Accordingly, the $(3N+3)$-vector $\vec{R}$ is given by
\begin{equation}\label{eq:R}
\vec{R}=\frac{EI}{1+\delta_{\alpha,3}\sigma}[0,0,0,\kappa_1(s_1),\kappa_2(s_1),\kappa_3(s_1),\kappa_1(s_2),\ldots,\kappa_3(s_N)]^\transpose\,,
\end{equation}
so that the local moment balance is expressed relative to the local director
basis. We remark that each of the quantities involved in the construction of
$\mathcal{B}$ and $\vec{R}$ are well-defined for a general filament in three
dimensions, given the local directors $\vec{d}_1$ and $\vec{d}_2$ and
computing the components of the twist vector as $\kappa_1 =
\vec{d}_3 \cdot \partial_s
\vec{d}_2$, $\kappa_2 =
\vec{d}_1 \cdot \partial_s \vec{d}_3$, and $\kappa_3 = \vec{d}_2 \cdot
\partial_s \vec{d}_1$. In terms of the discretised filament, these arclength
derivatives are approximated via finite differences in practice. Additionally,
we will proceed assuming that the filament is moment-free at the base, which
additionally enforces $\kappa_1(0)=\kappa_2(0)=\kappa_3(0)=0$.

\subsection{Coupling hydrodynamics}
We now relate the force density $\vec{f}$ acting on the fluid to the velocity
of each segment endpoint, utilising the commonly-applied method of resistive
force theory as introduced by \citet{Hancock1953,Gray1955} and adopted by
\citet{Moreau2018} for planar filaments, incurring typical errors logarithmic
in the aspect ratio of the filament. Here taking the radius of the filament to
be $\epsilon=10^{-2}L$, which more generally is assumed to be small in
comparison to the filament length, simple resistive force theory gives the
leading order relation between filament velocity and force density as
\begin{equation}
	f_t = -C_t u_t\,, \quad f_n = -C_n u_n\,.
\end{equation}
Here $f_t$ and $f_n$ denote the components of the force density tangential and
normal to the filament, with analogous definitions of $u_t$ and $u_n$. We will
utilise the expression of \citet{Gray1955}, with
\begin{equation}
	C_t = \frac{2\pi\mu}{\log\left(2L/\epsilon\right) - 0.5}\,, \quad C_n = \frac{4\pi\mu}{\log\left(2L/\epsilon\right) - 0.5}\,,
\end{equation}
where $\mu$ is the medium viscosity, noting the relation $C_n=2C_t$. We
approximate the local filament tangent at the segment endpoint $\vec{x}_i$ as
the average of $\vec{d}_3^{i-1}$ and $\vec{d}_3^i$ for $i=2,\ldots,N$, with
the tangent for $i=1$ and $i=N+1$ simply being taken as $\vec{d}_3^1$ and
$\vec{d}_3^{N}$ respectively. By linearity, and again assuming a
piecewise-linear force density along segments, we may write the coupling of
translational kinematics to hydrodynamics as
\begin{equation}\label{eq:fluid}
	\dot{\vec{X}} = A\vec{F}\,,
\end{equation}
where $A$ is a square matrix of dimension $3(N+1)\times3(N+1)$ and is a
function only of the segment endpoints $\vec{x}_i$. Of dimension $3(N+1)$, the
vector $\dot{\vec{X}}$ corresponds to the linear velocities of the segment
endpoints, and is constructed analogously to $\vec{F}$ with respect to the
laboratory frame. This relation results from the application of the no-slip
condition at the segment endpoints, coupling the filament to the surrounding
fluid.

In order to relate the rate of rotation of each segment to the viscous torque
$\vec{\tau}_i$ acting on it, we here consider an approximation of the finite
segment as an infinite rotating cylinder, associating the torque per unit
length on the $i$\textsuperscript{th} segment with the rotation $\omega_i$
about its local tangent $\vec{d}_3^i$ via the relation of \citet{Chwang1974}:
\begin{equation}\label{eq:rot}
	\vec{\tau}_i = 4\pi\mu\epsilon^2\omega_i\vec{d}_3^i\
\end{equation}
and in particular the $\epsilon^2$ scaling entails the torque per unit length
contributions are relatively small. Here we recall that $\mu$ is the viscosity
of the fluid medium, and $\epsilon$ is the radius of the filament. We may
write this relation as a linear operator on
$\vec{\omega}=[\omega_1,\ldots,\omega_N]^\transpose$, written simply as
$\vec{T}=\tilde{A}\vec{\omega}$. This crude approximation may readily be
substituted for non-local hydrodynamics via the method of regularised
Stokeslet segments, which will likely be a topic of future work. Similarly,
non-local hydrodynamics may be utilised in place of \cref{eq:fluid}, as used
for two-dimensional filament studies by \citet{Hall-McNair2019} and
\citet{Walker2019d}, the latter incorporating a planar no-slip boundary and
still yielding an explicit linear relation analogous to \cref{eq:fluid}.

Combining \cref{eq:BFR,eq:fluid,eq:rot} yields the linear system
\begin{equation}\label{eq:BAXR}
	-\mathcal{B}\left[\begin{array}{cc}
		A^{-1} & 0\\ 
		0 & \tilde{A}
	\end{array}\right]
	\left[\begin{array}{c}
	\dot{\vec{X}} \\
	\vec{\omega}
	\end{array}\right]=-\mathcal{B}\mathcal{A}\left[\begin{array}{c}
	\dot{\vec{X}} \\
	\vec{\omega}
	\end{array}\right]=\vec{R}\,,
\end{equation}
where $A$ is invertible and $\mathcal{A}$ is defined to be a block matrix of
dimension $(6N+3)\times(4N+3)$ with non-zero blocks $A^{-1}$ and $\tilde{A}$.

\subsection{Parameterisation}
We may parameterise the tangents $\vec{d}_3^i$ on each linear segment by the
Euler angles $\theta_i\in[0,\pi]$, $\phi_i\in(-\pi,\pi]$,
$\psi_i\in(-\pi,\pi]$ for $i=1,\ldots,N$ \cite{Antman2005}. With this
parameterisation we may make a choice of $\vec{d}_1$ and $\vec{d}_2$, taking
here the three orthonormal vectors to be
\begin{align}
	\vec{d}_1^i &= [-\sp\cpsi-\ct\cp\spsi,+\cp\cpsi-\ct\sp\spsi,\st\spsi]^\transpose\,, \label{eq:directors1}\\
	\vec{d}_2^i &= [+\sp\spsi-\ct\cp\cpsi,-\cp\spsi-\ct\sp\cpsi,\st\cpsi]^\transpose\,, \label{eq:directors2}\\
	\vec{d}_3^i &= [\st\cp,\st\sp,\ct]^\transpose\,,\label{eq:directors3}
\end{align}
written with respect to the laboratory frame and where
$\st{}\equiv\sin{\theta_i}$, $\ct{}\equiv\cos{\theta_i}$, and analogously for
$\sp{},\cp{},\spsi$ and $\cpsi$. From the directors we recover

\begin{equation}\label{eq:angles}
 	\theta_i = \arccos{\left(\vec{d}_3^i \cdot \vec{e}_z\right)}\,,\quad
 	\phi_i = \arctan{\left(\frac{\vec{d}_3^i \cdot \vec{e}_y}{\vec{d}_3^i \cdot \vec{e}_x}\right)}\,,\quad
 	\psi_i = \arctan{\left(\frac{\vec{d}_1^i \cdot \vec{e}_z}{\vec{d}_2^i \cdot \vec{e}_z}\right)}\,.
\end{equation}
As the discretised filament is piecewise linear, for $j=1,\ldots,N+1$ we may
write
\begin{equation}
	\vec{x}_j = \vec{x}_1 + \ds\sum\limits_{i=1}^{j-1}\vec{d}_3^i\,, \quad
	\dot{\vec{x}}_j = \dot{\vec{x}}_1 + \ds\sum\limits_{i=1}^{j-1}\dot{\vec{d}}_3^i\,.
\end{equation}
With $\vec{d}_3^i$ parameterised as above, we can thus express
$\dot{\vec{x}}_j$ as a linear combination of the derivatives of $\theta_i$ and
$\phi_i$ for $i=1,\ldots,j-1$, in addition to including the time derivative of
the base point $\vec{x}_1$. Hence we may write
\begin{gather}\label{eq:QTX}
	Q\dot{\vec{\Theta}} = \dot{\vec{X}}\,, \\
	\vec{\Theta}=[x_{1,x},x_{1,y},x_{1,z},\theta_1,\ldots,\theta_N,\phi_1,\ldots,\phi_N,\psi_1,\ldots,\psi_N]^\transpose\,,
\end{gather}
where $Q$ is a $3(N+1)\times3(N+1)$ matrix and $x_{1,x},x_{1,y,}x_{1,z}$ are
the components of $\vec{x}_j$ in the basis
$\{\vec{e}_x,\vec{e}_y,\vec{e}_z\}$. Explicitly, $Q$ may be constructed via
\begin{equation}
	\tilde{Q} = \left[
	\begin{array}{c|c|c|c}
	Q_{11} & Q_{12} & Q_{13} & \\
	\cline{1-3}
	Q_{21} & Q_{22} & Q_{23} & 0 \\
	\cline{1-3}
	Q_{31} & Q_{32} & Q_{33} \\	
	\end{array}\right]\,, \quad Q = [\tilde{Q}]_P\,,
\end{equation}
where the matrices $Q_{k1}$ are of dimension $(N+1)\times3$, with $Q_{k2}$ and
$Q_{k3}$ being of dimension $(N+1)\times N$, for $k=1,2,3$. In the definition
of $Q$, the subscript $P$ denotes that the $i$\textsuperscript{th} row of
$\tilde{Q}$ is permuted to the $P(i)$\textsuperscript{th} row of $Q$, where
\begin{equation}
	P(i)=\left\{\begin{array}{rl}
	3(i-1)+1\,, & i = 1,\ldots,N+1\,,\\
	3(i-N-2)+2\,, & i = N+2,\ldots,2N+2\,,\\
	3(i-2N-3)+3\,, & i = 2N+3,\ldots,3N+3\,.
\end{array}\right.
\end{equation}
This permutation of $\tilde{Q}$ allows us to define the sub-blocks simply,
given explicitly as
\begin{align*}
	Q_{k1}^{i,j} &=\left\{\begin{array}{rl}
		1\,, & j=k\,,\\
		0\,, & \text{otherwise}\,,
	\end{array}\right. \quad k=1,2,3\,,\\
	Q_{12}^{i,j} &=\ds\left\{\begin{array}{rl}
		+\cos{\theta_j}\cos{\phi_j}\,, & j<i\,,\\
		0\,, & j\geq i\,,
	\end{array}\right.\\
	Q_{13}^{i,j} &=\ds\left\{\begin{array}{rl}
		-\sin{\theta_j}\sin{\phi_j}\,, & j<i\,,\\
		0\,, & j\geq i\,,
	\end{array}\right.\\
	Q_{22}^{i,j} &=\ds\left\{\begin{array}{rl}
		+\cos{\theta_j}\sin{\phi_j}\,, & j<i\,,\\
		0\,, & j\geq i\,,
	\end{array}\right.\\
	Q_{23}^{i,j} &=\ds\left\{\begin{array}{rl}
		+\sin{\theta_j}\cos{\phi_j}\,, & j<i\,,\\
		0\,, & j\geq i\,,
	\end{array}\right.\\
	Q_{32}^{i,j} &=\ds\left\{\begin{array}{rl}
		-\sin{\theta_j}\,, & j<i\,,\\
		0\,, & j\geq i\,,
	\end{array}\right.\\
	Q_{33}^{i,j} &= 0\,.
\end{align*}

Further, in this parameterisation we may readily relate the local rate of
rotation about $\vec{d}_3^i$, denoted $\omega_i$ as in \cref{eq:rot}, to
$\theta,\phi,\psi$ and their time derivatives. Explicitly, this relationship
is $\omega=\cos({\theta})\dot{\phi}+\dot{\psi}$, and is notably linear in the
derivatives of the Euler angles. Thus, we form the composite matrix
\begin{equation}
	\mathcal{Q} = 
	\left[\begin{array}{c|cc}
		\multicolumn{3}{c}{Q} \\
		\hline
		\multicolumn{2}{c|}{0 \ \vline \ C} & I_N
	\end{array}\right]\,,
\end{equation}
where $I_N$ is the $N\times N$ identity matrix and the $N\times N$ matrix $C$
has diagonal elements $C_{i}=\cos({\theta_i})$ for $i=1,\ldots,N$, with all
other elements zero. The upper block, $Q$, maps the parameterisation into the
laboratory frame, whilst the lower blocks convert between the parameterisation
and the local rate of rotation about $\vec{d}_3$ as written in director basis.
The $(4N+3)\times3(N+1)$ matrix $\mathcal{Q}$ now encodes the expressions of
velocities and rotation rates in terms of the parameterisation, via
\begin{equation}\label{eq:QTXW}
	\mathcal{Q}\dot{\vec{\Theta}} = \left[\begin{array}{c}
	\dot{\vec{X}} \\
	\vec{\omega}
	\end{array}\right]\,,
\end{equation}
noting that the representation of $\vec{\omega}$ is relative to the director
basis, whilst the representation of $\dot{\vec{X}}$ is relative to the basis
of the laboratory frame. 

Having constructed $\mathcal{Q}$, we now combine \cref{eq:BAXR,eq:QTXW} to
give
\begin{equation}\label{eq:final_system}
	-\mathcal{B}\mathcal{A}\mathcal{Q}\dot{\vec{\Theta}} = \vec{R}\,,
\end{equation}
noting in particular that the matrix $\mathcal{B}\mathcal{A}\mathcal{Q}$ is
square and of dimension $(3N+3)\times(3N+3)$. Naively, this system of ordinary
differential equations can be readily solved numerically to give the evolution
of the filament in the surrounding fluid. However, the use of a single
parameterisation to describe the filament will in general lead to degeneracy
of the linear system and ill-defined derivatives in both space and time,
issues which we explore and resolve numerically in the subsequent sections.

\subsection{Coordinate singularities}
Consider a straight filament aligned with the $\vec{e}_z$ axis, with each of
the $\vec{d}_3^i=[0,0,1]^\transpose$ written in the laboratory frame. For this
filament $\theta_i=0$ for all $i$, whilst the $\phi_i$ are undetermined,
arbitrary and notably need not be the same on each segment. Were we to attempt
to formulate and solve the linear system of \cref{eq:final_system}, both
$\phi$ and its derivatives would be ill-defined, and correspondingly we would
be unable to solve the system for the filament dynamics, which are physically
trivial in this particular setup. In more generality, if a filament were to
have any segment pass through one of the poles $\theta=\{0,\pi\}$ of this
coordinate system, $\phi$ would be undetermined on the segment and arbitrary,
with attempts to solve our parameterised system of ordinary differential
equations failing. Further, were a segment to pass close to but not through a
pole, time derivatives of $\phi$ would necessarily become large, with $\phi$
well-defined but varyingly rapidly as the segment moves close to the pole of
the coordinate system. These large derivatives would artificially introduce
additional stiffness to the elastohydrodynamical problem, inherent only to the
parameterisation and not the underlying physics. This problem is well known
for Euler angle parameterisations, and is commonly referred to as the `gimbal
lock'. Analogous issues with arclength derivatives occur when considering
neighbouring segments, with the value of $\phi$ varying rapidly and
artificially between segments that reside near the pole of the coordinate
system. In this latter case however, our formulation of the
elastohydrodynamical problem circumvents the need for evaluation of $\phi_s$,
instead considering only derivatives of the smooth quantities
$\vec{d}_\alpha$, though we are not able to resolve issues with temporal
derivatives in the same way.

In order to avoid the numerical and theoretical problems associated with
singular points in the filament parameterisation, we exploit the finiteness of
the set of angles $\theta_i$ along with the independence of the underlying
elastohydrodynamical problem from the parameterisation. Throughout this work
we have assumed a fixed laboratory frame with basis
$\{\vec{e}_x,\vec{e}_y,\vec{e}_z\}$, present only so that vector quantities
may be written component-wise for convenience. Our choice of such a basis is
arbitrary, with the physical problem of filament motion being independent of
our selection of particular basis vectors. It is with respect to this basis
that we have defined the Euler angles $\theta,\phi,\psi$, from which the
aforementioned coordinate singularities appear if any of the $\theta_i$
approach zero or $\pi$. Thus, if one makes a choice of basis
$\{\vec{e}_x^\star,\vec{e}_y^\star,\vec{e}_z^\star\}$ such that the
corresponding Euler angles $\theta_i^\star$ are some $\delta$-neighbourhood
away from the poles of the new parameterisation, the system of ordinary
differential equations given in \cref{eq:final_system} may be readily solved,
at least initially. Should the solution in the new coordinate system approach
one of the new poles $\theta^\star=0,\pi$, a new basis can again be chosen,
and this process iterated until the filament motion has been captured over a
desired interval.

We note that for sufficiently small $\delta>0$ such a choice of basis
$\{\vec{e}_x^\star,\vec{e}_y^\star,\vec{e}_z^\star\}$ necessarily exists due
to the finiteness of the set of $\theta_i$, with $\delta$ in practice able to
be sufficiently large so as to limit the effects of coordinate singularities.
Thus, subject to reasonable assumptions of smoothness of the filament position
$\vec{x}$, such a process of repeatedly changing basis when necessary will
prevent issues associated with the parameterisation described above, and will
in practice enable the efficient simulation of filament motion without
introducing significant artificial stiffness or singularities.

\section{Implementation, verification and extensions}
\subsection{Selecting a new basis}

\begin{figure}[ht]
	\centering
	\subfloat[\label{fig:sphere:before}]{%
        \includegraphics[width=0.42\textwidth]{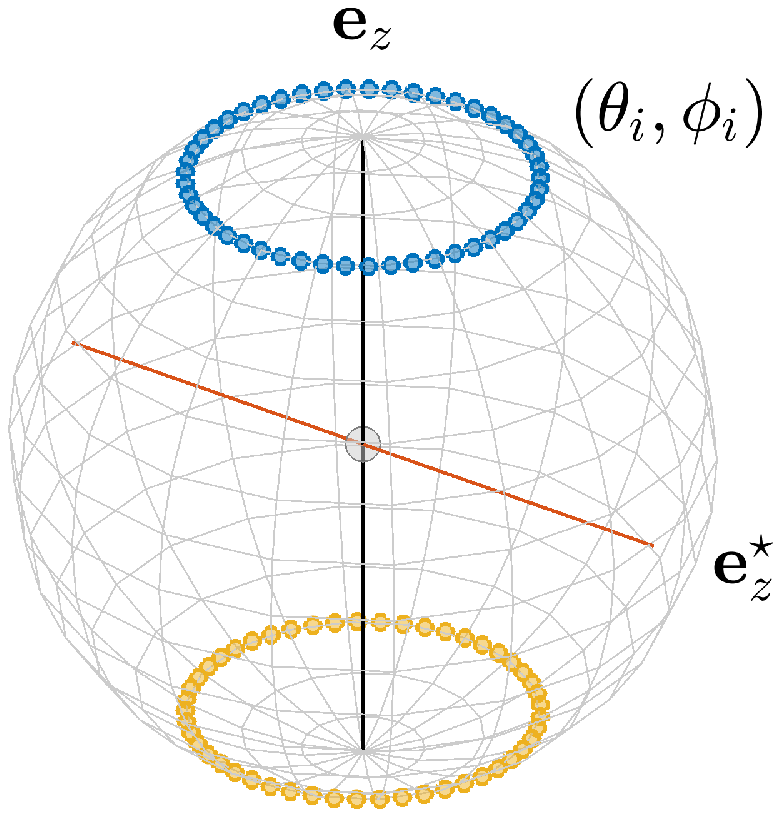}}
    \subfloat[\label{fig:sphere:after}]{%
        \includegraphics[width=0.42\textwidth]{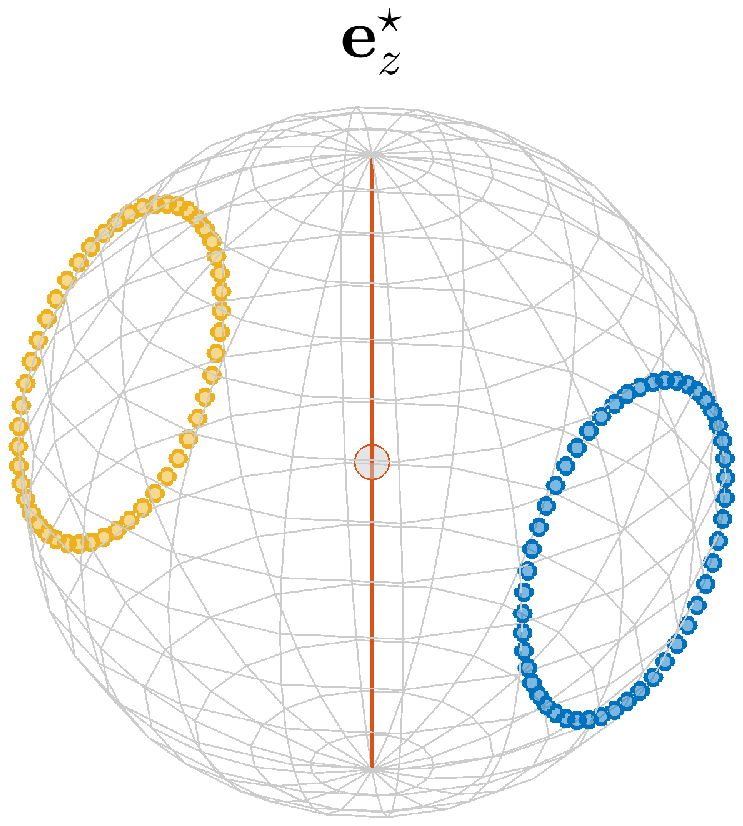}}
	\caption{An example choice of a new parameterisation in order to avoid
    coordinate singularities. Shown in blue are the points on the unit sphere
    corresponding to $(\theta_i,\phi_i)$, with their antipodes displayed in
    yellow. (a) Before a change of basis and subsequent reparameterisation, we
    see that these points are located near to the $\theta=0$ and $\theta=\pi$
    axes, which are shown as black vertical lines. A new potential location
    for the $\theta=0$ axis is shown in orange, selected so as to maximise the
    distance from the $(\theta_i,\phi_i)$ and their antipodes. (b) Following
    reparameterisation, the points and antipodes are located maximally away
    from the new axis. This example scenario corresponds to a helical filament
    initially with $\theta_i=\pi/6$ for $i=1,\ldots,50$.}
	\label{fig:sphere}
\end{figure}

Initially choosing an arbitrary laboratory basis
$\{\vec{e}_x,\vec{e}_y,\vec{e}_z\}$, the above formulation is implemented in
MATLAB\textsuperscript{\textregistered}, with the system of ordinary
differential equations (ODEs) of \cref{eq:final_system} being solved using the
inbuilt stiff ODE solver \texttt{ode15s}, as described in detail by
\citet{Shampine1997}. This standard variable-step, variable-order solver
allows for configurable error tolerances, typically set here at $10^{-5}$ for
absolute error and $10^{-4}$ for relative error, in general significantly
below the error associated with the piecewise-linear filament discretisation.
Derivatives with respect to arclength are approximated with fourth order
finite differences, with the resulting dynamics insensitive to this choice of
scheme. Initially and at each timestep, the values of $\theta_i$ are checked
to determine if they are within $\delta$ of a coordinate singularity,
typically with $\delta=\pi/50$. Should the parameterisation be approaching a
singularity, a new basis is chosen and the problem recast in this basis.

A natural method of selecting a new basis is perhaps to choose one uniformly
at random. Indeed, by considering the worst-case scenario of the $N$ tuples
$(\theta_i,\phi_i)$ uniformly and disjointly covering the surface of the unit
sphere, which together $\theta$ and $\phi$ parameterise, the probability that
any random basis results in a scenario with
$\min_i{\{\theta_i,\pi-\theta_i\}}<\delta$ is given by $2N\sin^2(\delta/2)$, a
consequence of elementary geometry. With this quantity being significantly
less than unity for a wide range of $N$ with $\delta$ large enough to avoid
severe artificial numerical stiffness, as discussed above, a practical
implementation for the simulation of filament elastohydrodynamics as
formulated above may simply select a new basis randomly, repeating until a
suitable basis is found. With $\delta=\pi/50$ and $N=50$, the probability of
rejecting a candidate new basis is bounded above by $10\%$, thus in practice
one should expect to find an appropriate basis within few iterations of the
proposed procedure.

Alternatively, and as we will do throughout this work, one may instead proceed
in a deterministic manner, selecting a near-optimal basis from knowledge of
the existing parameterisation. Given the set of parameters $\theta_i$ and
$\phi_i$, we may choose a $\hat{\theta}\in[0,\pi]$ and $\hat{\phi}\in[0,2\pi)$
so as to maximise the distance of $(\hat{\theta},\hat{\phi})$ from each of the
$(\theta_i,\phi_i)$ and their antipodes. In practice, an approximate solution
to this problem is attained by selecting $(\hat{\theta},\hat{\phi})$ from a
selection of preset test points in order to maximise the distance from the
$(\theta_i,\phi_i)$, where distance is measured on the surface of the unit
sphere that $\theta$ and $\phi$ naturally parameterise, as shown in
\cref{fig:sphere}. It should be noted that this process of selection impacts
negligibly on computational efficiency with 10,000 test points. With these
choices of $\hat{\theta}$ and $\hat{\phi}$, we form a new basis by mapping the
original basis vector $\vec{e}_z$ to the vector $\vec{e}_z^{\star}$, given
explicitly by
\begin{equation}
	\vec{e}_z^{\star} =
	[\sin{\hat{\theta}}\cos{\hat{\phi}},\sin{\hat{\theta}}\sin{\hat{\phi}},\cos{\hat{\theta}}]^\transpose\,.
\end{equation}
Choosing the other members of the orthonormal basis
$\vec{e}_x^{\star},\vec{e}_y^{\star}$ arbitrarily, expressed in this new basis
the accompanying filament parameterisation will be removed from any coordinate
singularities by construction, as exemplified in \cref{fig:sphere:after}.

\subsection{Validation}
In what follows we validate the presented methodology against known filament
behaviours and a sample three dimensional simulation via an existing
methodology. Initial configurations and parameter values for each can be found
in the \cref{app:params}, with behaviours qualitatively independent of these
parameter choices and filament setups.

\subsubsection{Relaxation of a planar filament}
\begin{figure}
	\centering
	\begin{overpic}[percent,width=\textwidth]{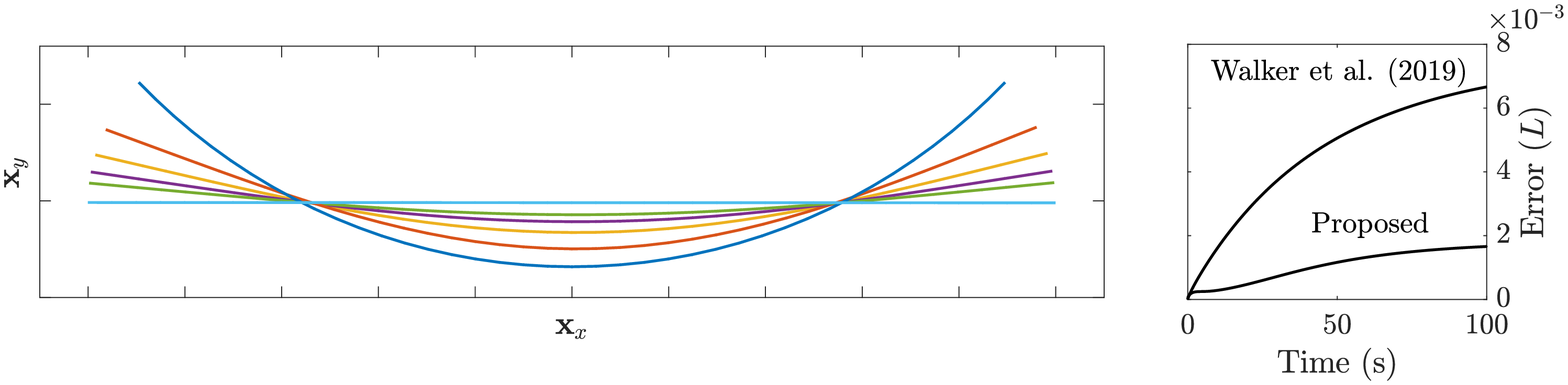}
	\put(-1,22){(a)}
	\put(72,22){(b)}
	\end{overpic}
	\caption{The two-dimensional relaxation of a symmetric planar filament,
simulated with $N=40$ segments. (a) Symmetry is preserved throughout the
dynamics, with the relaxation in qualitative agreement with that used for
verification in the two-dimensional works of
\citet{Moreau2018,Hall-McNair2019,Walker2019d}. (b) Translation of the centre
of mass throughout the motion, analytically zero, is captured numerically with
errors on the order of $10^{-3}L$ by the proposed methodology, notably the
same order of magnitude as that attained with the two-dimensional methodology
of \citet{Walker2019d}. Axes $\vec{x}_x$ and $\vec{x}_y$
correspond to the unit vectors $\vec{e}_x$ and $\vec{e}_y$ respectively.}
	\label{fig:2d-relax}
\end{figure}

Noting there is no analytical test solution for the dynamics of a fully 3D
Kirchhoff rod in a viscous fluid, to the best of our knowledge, we consider
validations by comparison with numerical studies in the literature, though we
additionally utilise invariance of the centre of mass as a gold standard
below, where applicable. Firstly, we validate the presented approach by
considering the problem of filament relaxation in two dimensions, a natural
and well-studied subset of the three dimensional framework. We consider the
simple case of a symmetric curved filament in the $\vec{e}_x\vec{e}_y$ plane,
which will provide a test of symmetry preservation, integrated moment balance,
and hydrodynamics via qualitative comparisons with the earlier works of
\citet{Moreau2018} and \citet{Hall-McNair2019}. Simulating the relaxation of
such a filament to a straight equilibrium condition with $N=40$ segments takes
less than \SI{2}{\s} on modest hardware
(Intel\textsuperscript{\textregistered} Core\texttrademark\ i7-6920HQ CPU),
from which we immediately see retained the computational efficiency of the
framework of \citeauthor{Moreau2018} which this work generalises. Present
throughout the computed motion is the left-right symmetry of the initial
condition, with the filament shape evolving smoothly even with a small number
of segments, as shown in \cref{fig:2d-relax}. Further, the centre of mass,
which in exact calculation would be fixed in space due to the overall
force-free condition on the filament, is captured numerically with errors on
the order of $10^{-3}L$, demonstrating very good quantitative satisfaction of
this condition. Of note, in \cref{fig:2d-relax}b we have verified that this
error is of the same order of magnitude as that generated by the
two-dimensional methodology of \citet{Walker2019d}.

\subsubsection{Planar bending of a filament in shear flow}
Further, whilst the above is reassuring and serves to validate a subset of the
implementation, we note from \cref{eq:directors1,eq:directors2} that motion
cast in the $\vec{e}_x\vec{e}_y$ plane of the laboratory frame may often
render the evolution of one of the directors $\vec{d}_1,\vec{d}_2$ trivial. In
order to avoid this we may consider planar problems in slightly more
generality, posing a problem that is planar though not aligned with the
$\vec{e}_x\vec{e}_y$ plane. As an exemplar such problem we attempt to recreate
a typical but complex behaviour of a flexible filament in a shear flow, that
of the \emph{J-shape} and \emph{U-turn} \cite{Liu2018}, aligning both the
filament and the background flow in a plane spanned by $\vec{e}_y$ and
$\vec{e}_x+\vec{e}_z$. In order to ensure the absence of alignment of the
parameterisation with the $\vec{e}_x\vec{e}_y$ plane, we disable the adaptive
system of basis selection for the purpose of this example, and simulate the
motion of a filament in a shear flow. The background flow with velocity
$\vec{u}_b$ and vorticity $\vec{\Omega}$ is incorporated into the framework
via the transformation $\vec{u}\mapsto\vec{u}-\vec{u}_b$, $\omega_i \mapsto
\omega_i - \vec{\Omega}\cdot\vec{d}_3^i/2$, yielding the modified system
\begin{equation}
	-\mathcal{B}\mathcal{A}\mathcal{Q}\dot{\vec{\Theta}} = \vec{R} -
	\mathcal{B}\mathcal{A}\vec{U}_b
\end{equation}
for a vector $\hat{\vec{U}_b}$ of background flows and vorticities evaluated
at segment endpoints and midpoints, respectively. Details of the flow field
and initial setup are given in the \cref{app:params}.

\begin{figure}
	\centering
	\begin{overpic}[permil,width=0.6\textwidth]{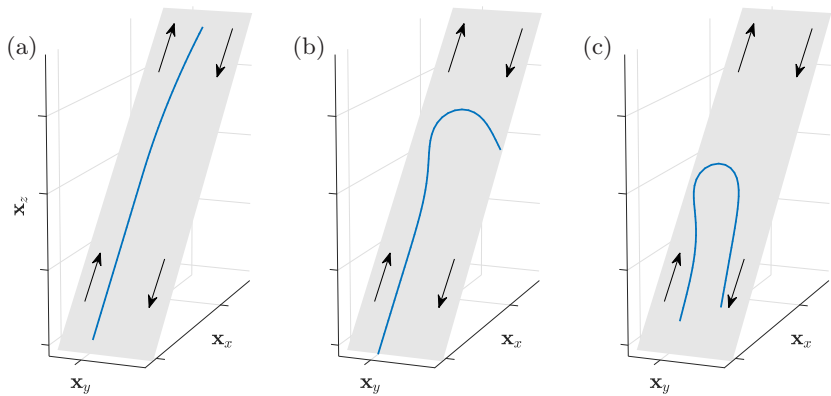}
		\put(-10,420){(a)}
		\put(340,420){(b)}
		\put(695,420){(c)}
	\end{overpic}
	\caption{The planar evolution of a filament in shear, exhibiting a rich
and well-documented dynamics. (a) Having aligned both the filament and shear
flow in a plane not parallel to the $\vec{e}_x\vec{e}_y$ plane, with the
directors non-trivial in this setup, we simulate the motion of the filament
through two distinct morphological transitions: (b) the characteristic
\emph{J-shape} and (c) the subsequent \emph{U-turn} (cf.~ \citet[Fig. 1,
Movies S5, S6]{Liu2018}). We note that the choosing of an improved basis for
computation has been disabled for this example, and yields a twofold increase
in computational efficiency if enabled. Arrows indicate the direction of the
background shear flow. Axes $\vec{x}_x,\vec{x}_y,\vec{x}_z$ correspond to the
unit vectors $\vec{e}_x,\vec{e}_y,\vec{e}_z$, respectively. The plane
containing the filament and the shear flow is shown in grey.}
	\label{fig:filament_in_shear}
\end{figure}

Adopting the timescale $T$ to be the
inverse of the shearing rate of the flow, we consider a parameter regime in
which one would expect to see to formation of J-shape and subsequently a
U-turn, defined by their characteristic morphologies in \citet[Figure 1,
Movies S5, S6]{Liu2018} and from which an appropriate parameter choice is
obtained. Notably, the impact of thermal noise perturbations is not considered
here, in contrast to \citet{Liu2018}, preventing a quantitative comparison. In
\cref{fig:filament_in_shear} we present the initial, J-shape, and U-turn
configurations of the filament as simulated via the proposed methodology, with
computation requiring approximately \SI{20}{\s} with $N=50$ segments. The
simulated filament shapes are in qualitative agreement with those shown in
\citet[Figure 1]{Liu2018}, and serve as further validation of the
coarse-grained methodology. Notably, enabling the described method of basis
selection effectively casts this problem in the $\vec{e}_x\vec{e}_y$ plane,
affording a twofold increase in computational efficiency and highlighting the
benefits of adaptive reparameterisation.

\subsubsection{Relaxation of a non-planar filament} 
\begin{figure}
\centering
\begin{overpic}[percent,width=0.8\textwidth]{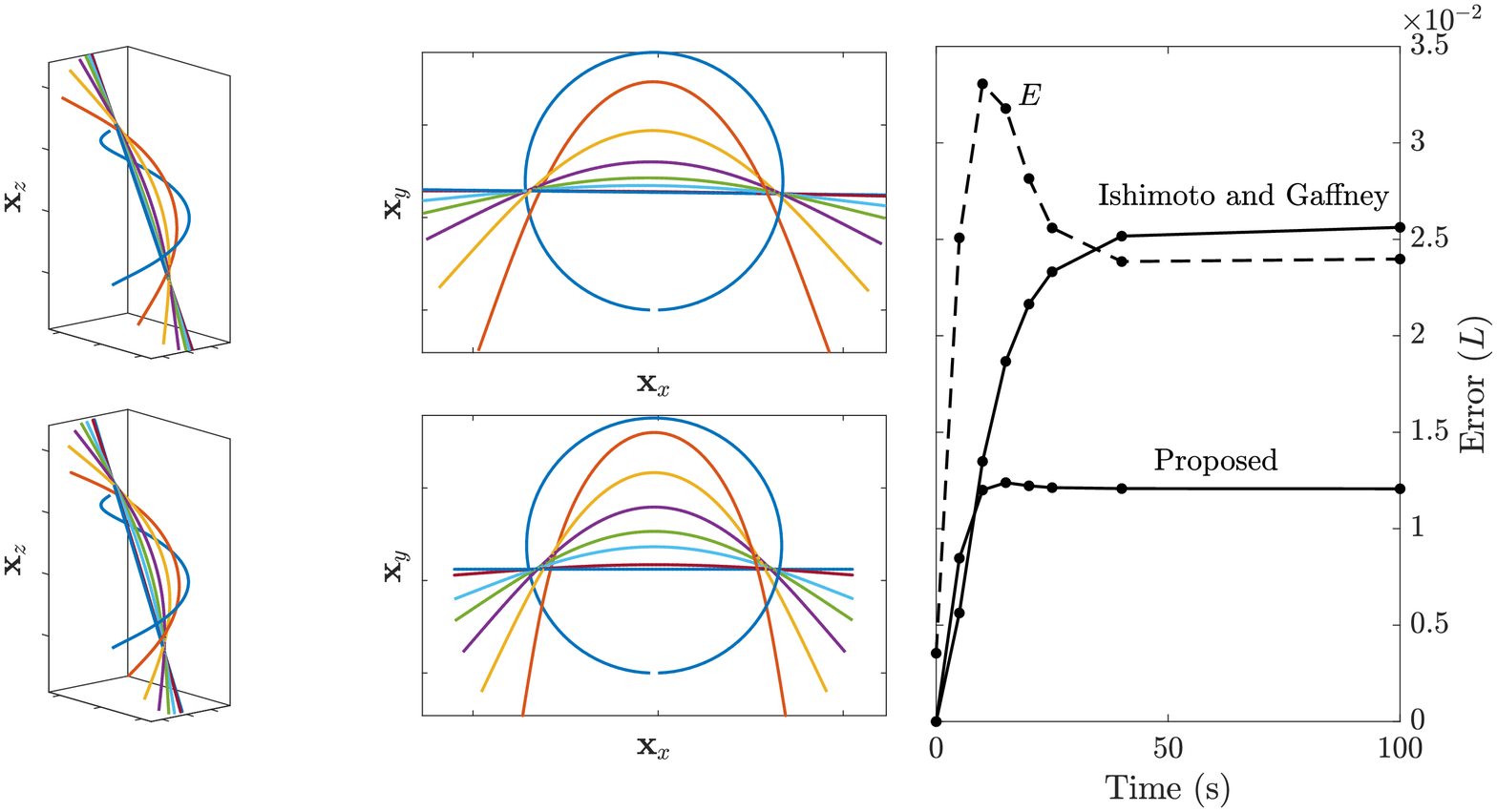}
\put(-1,53){(a)}
\put(24,53){(b)}
\put(-1,28){(c)}
\put(24,28){(d)}
\put(59,53){(e)}
\end{overpic}
\caption{The relaxation of a filament in three dimensions. Shown are the
results of simulating this motion via (a,b) the proposed framework and (c,d)
the methodology presented in \citet{Ishimoto2018a}, in turn heavily based on
the work of \citet{Olson2013} shown from multiple perspectives at multiple
timepoints. In both cases we observe relaxation from a non-planar, helical
configuration to a straight filament, and good agreement between the two
computed solutions, in particular given the logarithmic accuracy of resistive
force theories. (e) A quantitative comparison between the two frameworks, with
solid lines showing the Euclidean distance of the filament centre of mass from
its initial location, analytically zero by the force-free condition though
here on the order of $10^{-2}L$. The dashed curve quantifies the error between
the computed solutions, denoted $E$, with the square of this error defined as
$E^2 = \int \norm{\vec{x}_P - \vec{x}_{IG}}_2^2\intd{s} /L$, where $\vec{x}_P$
and $\vec{x}_{IG}$ are the locations of the filament centreline as computed by
the proposed methodology and that used by \citeauthor{Ishimoto2018a},
respectively. Computation with the presented methodology took approximately
\SI{30}{\s} on a modest laptop computer, in comparison to the multiple hours
required on sophisticated cluster hardware for the methodology of
\citeauthor{Ishimoto2018a}. Here we have simulated filament motion with
$N=100$ segments. Axes $\vec{x}_x,\vec{x}_y,\vec{x}_z$ correspond to the unit
vectors $\vec{e}_x,\vec{e}_y,\vec{e}_z$, respectively.}
\label{fig:relax}
\end{figure}

Finally, we consider truly
non-planar relaxation of filaments in three dimensions. Typical simulations of
such a relaxation with $N=50$ segments have a runtime of approximately
10\si{\s} on the modest hardware described above, often requiring at most one
choice of basis though naturally problem dependent, and provide reasonable
accuracy. Thus, even when considering inherently three-dimensional problems we
see retained in this methodology the low computational cost of the formulation
of \citet{Moreau2018}, representing significant improvements in computational
efficiency over recent studies in three dimensions
\cite{Olson2013,Simons2015,Ishimoto2018a}. This is particularly evident when
directly comparing the presented coarse-grained methodology with the results
of \citet{Ishimoto2018a}, considering in this case the relaxation of a helical
configuration to a straight equilibrium. A side-by-side comparison of the
relaxation dynamics as computed by the proposed methodology and that presented
in \citeauthor{Ishimoto2018a} is shown in \cref{fig:relax}, noting that the
work of \citet{Ishimoto2018a} considers an actively driven nearly inextensible
filament, of which relaxation dynamics are a natural subset. \Cref{fig:relax}
highlights good agreement between methodologies that is in line with the level
of accuracy typically afforded by resistive force theories used here,
recalling errors logarithmic in the filament aspect ratio. \Cref{fig:relax}e
shows a quantitative evaluation of the computed solutions, with the deviation
of the filament centre of mass from the initial condition shown as solid black
curves for each methodology. With the force-free condition implying that the
filament centre of mass should not move throughout the relaxation dynamics,
this measured deviation serves as an assessment of the accuracy of both
frameworks, with each exhibiting variation only on the order of $10^{-2}L$.
Also shown in \cref{fig:relax}e is a dimensional measure of the difference
between the two computed solutions, here denoted $E(t)$ and defined as the
non-negative root of
\begin{equation}
	E^2(t) = \frac{1}{L}\int\limits_0^L \norm{\vec{x}_P(s,t) -
	\vec{x}_{IG}(s,t)}_2^2\intd{s}\,,
\end{equation}
where $\vec{x}_P(s,t)$ and $\vec{x}_{IG}(s,t)$ denote the filament centreline
as computed by the proposed methodology and that used by
\citeauthor{Ishimoto2018a}, respectively. Numerically approximating this
integral with quadrature and noting that this error is consistently on the
order of $10^{-2}L$ throughout the relaxation, we see evidenced good
quantitative agreement between the two frameworks, thus validating the
proposed methodology. We also remark that the solution of
\citet{Ishimoto2018a} does not perfectly satisfy filament inextensibility,
with variation in total length of approximately 1\%, which may have some
impact on the computed dynamics. Thus, when computing $E(t)$ as defined above,
we treat $s\in[0,L]$ as a material parameter, with the differences in position
$\vec{x}_P(s,t) - \vec{x}_{IG}(s,t)$ therefore capturing discrepancies between
the simulated locations of the material point with undeformed arclength $s$ at
time $t$, with $s$ not necessarily equal to the deformed arclength in the
solution of \citet{Ishimoto2018a}.

\begin{figure}
\centering
\includegraphics[width=0.4\textwidth]{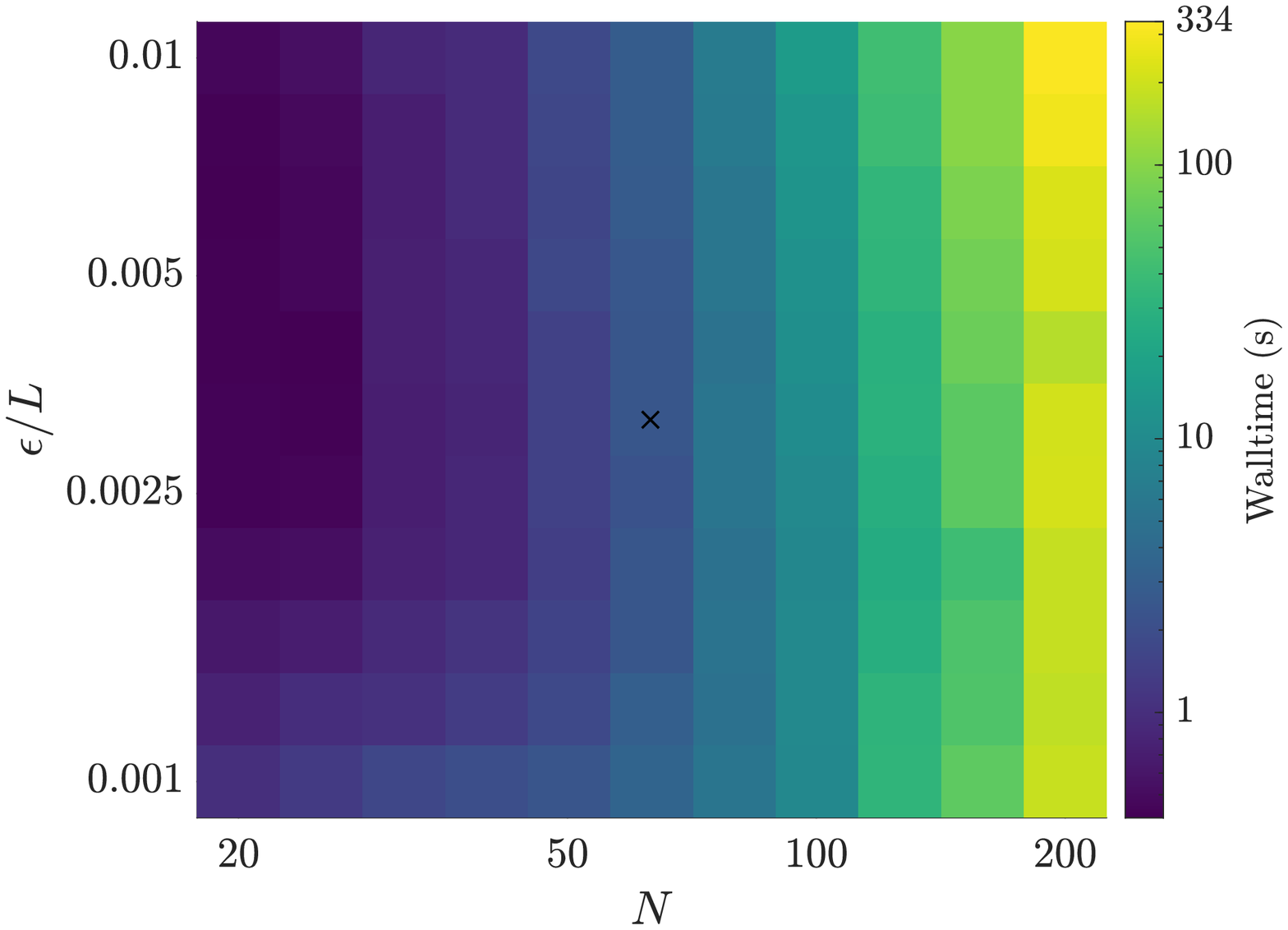}
\caption{Walltimes associated with the 3D filament relaxation of
\cref{fig:relax} for various choices of the parameters $\epsilon$ and $N$.
With $N=200$ corresponding to a very fine discretisation of the filament, we
observe a maximum walltime of under 6 minutes on a modest laptop computer,
approximately 25x faster than the similar computations of
\citet{Ishimoto2018a} on sophisticated cluster hardware, albeit for different
parameter values. There is a strong dependence of the walltime on the level of
discretisation, $N$, as expected. Here, the absolute and relative error
tolerances of the ODE solver are set to $10^{-5}$ and $10^{-4}$, respectively.
All axes, including the colour axis, are logarithmically scaled for visual
clarity. For reference, the parameter combination signified by a cross
corresponds to a walltime of
\SI{2.5}{\second}, with $N=63$ and $\epsilon/L\approx\SI{3.2e-3}{}$.
\label{fig:timings}}
\end{figure}

In contrast to the agreement between solutions, there is a stark difference
between the associated time required for computation. Taking $N=100$ segments
and computing until relaxation, the coarse-grained framework calculated the
solution in approximately \SI{30}{\s} on personal computing hardware with ODE
error tolerances of $10^{-5}$, whilst the computations utilising the
implementation of \citeauthor{Ishimoto2018a} required \SI{2.5}{\hour} on a
high performance computing cluster. A thorough investigation of the time
required for computation with the presented methodology for various choices of
the parameters $\epsilon$ and $N$ is showcased in \cref{fig:timings}, from
which we note the remarkable performance of this simple implementation across
parameter regimes, with the walltime naturally dependent on the discretisation
parameter $N$.

\subsection{Model extensions}
\subsubsection{Intrinsic curvature}
\label{sec:intrinsic_curvature}
In order to showcase the versatility of the presented framework, we
demonstrate its simple extension to filaments with non-zero intrinsic or
reference curvature, which can exhibit complex buckling behaviours
\cite{Lim2010}. Recalling the constitutive relation of
\cref{eq:constitutive_moments}, the effect of an intrinsic curvature
$\vec{\kappa}^0$ is to alter the bending moments, yielding the modified
constitutive relation
\begin{equation}\label{eq:intrinsic_curvature_constitutive}
	\vec{m} = EI\left([\kappa_1-\kappa_1^0]\vec{d}_1 + [\kappa_2-\kappa_2^0]\vec{d}_2 + \frac{1}{1+\sigma}[\kappa_3-\kappa_3^0]\vec{d}_3\right)\,,
\end{equation}
where we have written
$\vec{\kappa}^0=\sum_{\alpha}\kappa_{\alpha}^0\vec{d}_{\alpha}$ in the local
director basis. Notably, the reference curvature can plausibly depend on a
variety of quantities, including time, arclength, and spatial position, with
an example being the modelling of an internally driven filament by a
time-dependent intrinsic curvature \cite{Schoeller2019,Olson2013}.

\begin{figure}
	\centering
	\includegraphics[width=0.8\textwidth]{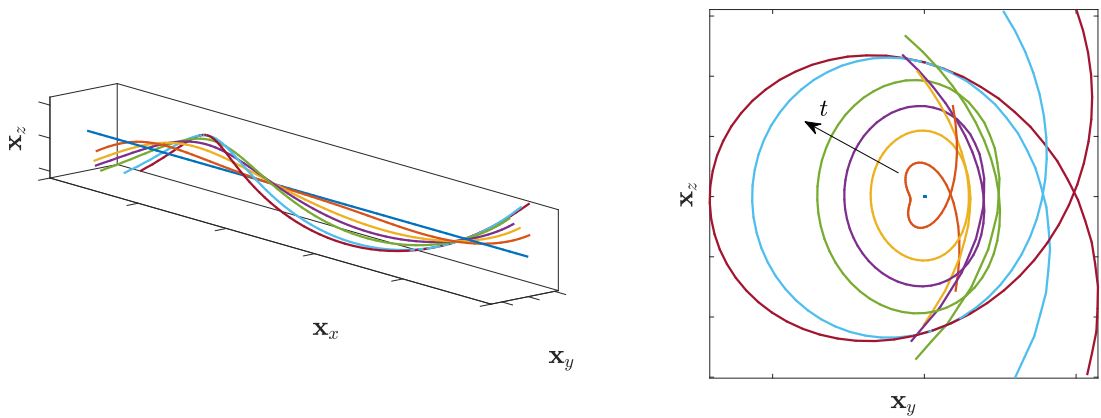}
	\caption{The relaxation of a straight filament with non-zero intrinsic
	curvature to a helical configuration, with reference curvature specified
	as $\vec{\kappa}^0=\pi\vec{d}_2 + 2\pi\vec{d}_3$. Having taken $N=70$
	segments and $\epsilon=10^{-2}L$, we observe the smooth relaxation of the
	filament away from its initial straight configuration, computed in
    \SI{15}{\second} on a modest laptop computer. Axes
    $\vec{x}_x,\vec{x}_y,\vec{x}_z$ correspond to the unit vectors
    $\vec{e}_x,\vec{e}_y,\vec{e}_z$, respectively. \label{fig:intrinsic}}
\end{figure}

Practically, the inclusion of such an intrinsic curvature amounts to a simple
subtraction of the reference curvature from the computed components of
$\vec{\kappa}$ at each instant, with $\vec{R}$ as given in \cref{eq:R} being
modified accordingly. Doing so, we simulate the relaxation of a straight
filament with a non-zero intrinsic curvature, with the reference curvature
explicitly given by $\vec{\kappa}^0=\pi\vec{d}_2 + 2\pi\vec{d}_3$
corresponding to a helical configuration. As shown in \cref{fig:intrinsic},
the filament indeed relaxes to a helix, with the walltime of this simulation
being \SI{15}{\second} on a laptop computer, having taken $\epsilon=10^{-2}L$
and $N=70$ segments, noting that the filament shape has been sufficiently
resolved with this discretisation.

\subsubsection{Clamped and internally driven filaments}
\label{sec:clamped_active}
Finally, we consider the simple extensions to both clamped filaments and to
those with actively generated internal moments, the latter being akin to the
active beating of biological cilia and flagella \cite{Gadelha2010}. For time
and arclength-dependent active moment density $\vec{m}^a$, its inclusion into
the presented framework acts to modify the moment balance of
\cref{eq:pointwise_moment_balance} to
\begin{equation}
    \vec{m}_s + \vec{x}_s\cross\vec{n} - \vec{\tau} + \vec{m}^a = \vec{0}\,.
\end{equation}
Repeating the integration of the pointwise moment balance from $s=s_i$ to
$s=L$ as in \cref{sec:methods} leads to a modified form of
\cref{eq:final_moment_balance}, explicitly given as
\begin{equation}
    -\vec{d}_{\alpha}^i \cdot\left(\vec{I}_i^f + \vec{I}_i^{\tau}\right) = \frac{EI}{1+\delta_{\alpha,3}\sigma}\kappa_{\alpha}(s_i) - \vec{d}_{\alpha}^i \cdot \vec{I}_i^a\,,
\end{equation}
where the integrated active moment density is written as
\begin{equation}
    \vec{I}_i^a = \int\limits_{s_i}^L\vec{m}^a\intd{\tilde{s}}\,.
\end{equation}
For a given active moment density, assumed to be integrable, $\vec{I}_i^a$ may
be readily computed either analytically or numerically, with its components in
the local $\vec{d}_{\alpha}^i$ direction then modifying $\vec{R}$ from
\cref{eq:R} accordingly.

\begin{figure}
    \centering
    \includegraphics[width=0.8\textwidth]{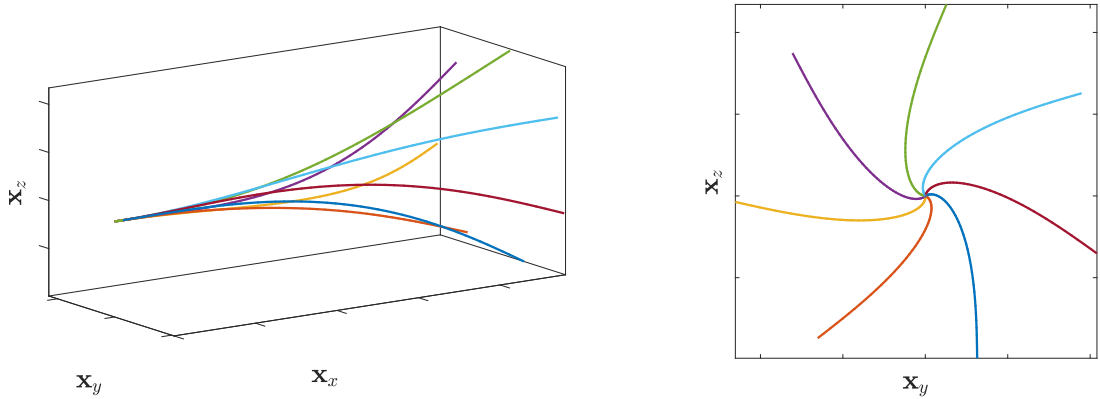}
    \caption{The 3D beating of a clamped filament driven by prescribed active internal
    moments. Having specified a travelling wave of out-of-phase sinusoidal
    active moments $\vec{m}^a = 5\sin(s-t)\vec{d}_1 + 5\cos(s-t)\vec{d}_2$, an
    initially straight filament deforms to a periodic driven motion, with the
    tip of the filament following a circular path. Here we have taken
    $\epsilon=10^{-2}L$ and $N=50$, noting that the filament shape has been
    resolved smoothly with this level of discretisation. Computation required
    approximately
    \SI{8}{\second} on modest hardware, simulating up until $t=8\pi$. Axes
    $\vec{x}_x,\vec{x}_y,\vec{x}_z$ correspond to the unit vectors
    $\vec{e}_x,\vec{e}_y,\vec{e}_z$, respectively.\label{fig:clamped_active}}
\end{figure}

Clamping the filament at the base is somewhat simpler, in that the overall
force and moment balance conditions on the filament are merely replaced by
enforcing no motion or rotation at the base. These conditions may be stated
concisely as
\begin{equation}
    \dot{\vec{x}}(0) = \vec{0}\,, \quad \dot{\theta_1} = \dot{\phi_1} = \dot{\psi_1} = 0\,,
\end{equation}
supplanting the force and moment-free conditions of
\cref{eq:total_force_balance,eq:total_moment_balance}, having taken $s=0$ in
the latter. Implementing these minor modifications, as an example we specify a
travelling wave of internal moment given by $\vec{m}^a = 5\sin(s-t)\vec{d}_1 +
5\cos(s-t)\vec{d}_2$ and simulate the active motion of a clamped filament,
taking $N=50$ and $\epsilon = 10^{-2}L$. Snapshots of this eventually periodic
motion are shown in \cref{fig:clamped_active}, with the motion simulated up
until $t=8\pi$ from a straight initial configuration and with a walltime of
around
\SI{8}{\second}.

\section{Discussion}
In this work we have seen that the motion of inextensible unshearable
filaments in three dimensions can be concisely described by a coarse-grained
framework, building upon the principles of \citet{Moreau2018} in order to
minimise numerical stiffness associated with the governing equations of
elastohydrodynamics. This representation was readily implemented via an Euler
angle parameterisation, with adaptive basis selection and reparameterisation
overcoming coordinate singularities associated with a fixed representation of
the dynamics. We have further demonstrated the efficacy of the proposed
deterministic method for basis selection by explicit simulation of non-planar
filament dynamics, which is able to afford reductions in numerical stiffness
even when separated from singularities of the parameterisation. The presented
framework retains the flexibility and extensibility of the formulations of
\citet{Moreau2018,Hall-McNair2019,Walker2019d}, with background flows, active
moment generation, body forces and other effects or constraints being simple
to include in this representation. The simplicity of these possible extensions
speaks to the broad utility of the proposed approach, with potential for use
in the simulation of both single and multiple filamentous bodies in fluid
under a wide variety of circumstances and conditions.

In formulating our methodology we have made the simplifying assumption of
coupling fluid dynamics to forces via resistive force theory, which is
well-known to incur errors logarithmic in the filament aspect ratio, though
variations remain in widespread use
\cite{Lauga2006,Gadelha2010,Sznitman2010,Curtis2012,Schulman2014,Utada2014,Moreau2018}.
Resistive force theories additionally suffer from locality, in that portions
of the filament do not directly interact with one another through the fluid. A
natural development of the presented approach would therefore be the inclusion
of non-local hydrodynamics, perhaps via the regularised Stokeslet segment
methodology of \citet{Cortez2001} as included in the work of
\citet{Walker2019d}, or lightweight singular slender body theories such as
those of \citet{Johnson1980,Tornberg2004,Walker2020b}. Such improvements may
also include the consideration of confined geometries, with motion in a half
space of particular pertinence to typical microscopy of flagellated organisms.
Despite the many possible directions for hydrodynamic refinement, we note that
the linearity of Stokes flows necessitates that any relations between forces
and velocities be linear, with explicit formulations simply giving rise to
modified linear operators $\mathcal{A}$ that may be readily inserted into the
described framework.

In the wider context of methods for soft filament simulation,
the scope of which is illustrated by the generality of the approach of
\citet{Gazzola2018}, the proposed framework enables rapid simulation of the
subclass of purely inextensible filaments. This modeling assumption has been
demonstrated to be a valid approximation of nearly inextensible filaments in a
variety of contexts \cite{Jabbarzadeh2020}, affording greatly improved
computational efficiency over previous approaches that we have compounded
here, albeit with reduced hydrodynamic fidelity. However, we expect that the
addition of improved hydrodynamics will have minimal impact on the
computational efficiency of the presented methodology, with this efficiency
not being derived from our use of simple resistive force theory, as noted by
\citet{Walker2019d,Hall-McNair2019} for their non-local refinements of the 2D
theory of \citet{Moreau2018}.

In summary, we have presented, verified and exemplified a novel
framework for the rapid simulation of inextensible, unshearable filaments in a
viscous fluid at zero Reynolds number. Despite the improved generality of this
methodology over existing two-dimensional approaches, we have retained the
computational efficiency and simplicity of the work of \citet{Moreau2018},
affording significant extensibility and thus facilitating a vast range of
previously unrealisable biological and biophysical studies into filament
dynamics on the microscale.

\section*{Acknowledgements}
We are grateful to Prof.\ Derek Moulton for discussions on elastic filaments,
and to Prof.\ David Smith for discussions on basis rotation. B.J.W.\ is
supported by the UK Engineering and Physical Sciences Research Council
(EPSRC), grant EP/N509711/1. K.I.\ acknowledges MEXT Leading Initiative for
Excellent Young Researchers (LEADER), JSPS KAKENHI for Young Researcher
(JP18K13456), and JST, PRESTO Grant Number JPMJPR1921, Japan. Elements of the
simulations were performed using the cluster computing system within the
Research Institute for Mathematical Sciences (RIMS), Kyoto University.

The computer code used and generated in this work is freely available from
\url{https://gitlab.com/bjwalker/3d-filaments.git}.

\appendix*
\section{Parameters and initial conditions}\label{app:params}
We nondimensionalise the system \cref{eq:final_system} as in
\citet{Walker2019d}, resulting in a dimensionless system of the form
\begin{equation}\label{eq:dimless}
	-E_h\hat{\mathcal{B}}\hat{\mathcal{A}}\hat{\mathcal{Q}}\dot{\hat{\vec{\Theta}}} = \hat{\vec{R}}\,, \quad E_h = \frac{8\pi\mu L^4}{EI\,T}
\end{equation}
for timescale $T$ and elastohydrodynamic number $E_h$, where dimensionless
counterparts are given by
\begin{equation}
	\mathcal{B} = L^2 \hat{\mathcal{B}}\,, \quad \mathcal{A} = \frac{1}{8\pi\mu}\hat{\mathcal{A}}\,, \quad \mathcal{Q}\dot{\vec{\theta}} = \frac{L}{T}\hat{\mathcal{Q}}\dot{\hat{\vec{\theta}}}\,, \quad \vec{R} = \frac{EI}{L}\hat{\vec{R}}\,,
\end{equation}
having multiplied the force balance equations by $\ds$ to unify dimensions.
All examples begin with the filament base, $\vec{x}_1$, coincident with the
origin of the laboratory frame, and we set Poisson's ratio to zero, i.e.~
$\sigma=0$ throughout though note a lack of sensitivity of results to this
choice.

\subsection{Relaxation of a planar filament}
In simulating the relaxation of a filament in the $\vec{e}_x\vec{e}_y$ plane,
we impose the initial filament shape as
\begin{equation}
	\theta_i=\pi/2 \,, \quad \phi_i = \frac{\pi}{2}\left(\frac{i-1}{N-1} - \frac{1}{2}\right)\,, \quad \psi_i=0\,, \quad \text{for }i = 1,\ldots,N
\end{equation}
with respect to standard laboratory Euler angles. Filament motion is simulated
with $E_h\approx \SI{1.6e5}{}$, though this choice is arbitrary given the
invariance of the dynamics to rescalings in time (assuming that the rescaling
is not so extreme as to break the inertialess assumption).

\subsection{Planar bending of a filament in shear flow}
In order to generate the characteristic behaviours of the \emph{J-shape} and
\emph{U-turn} we initialise the filament via
\begin{equation}
	\theta_i=\pi/4 \,, \quad \phi_i = -\frac{\pi}{12}\frac{i-1}{N-1}\,, \quad \psi_i=0\,, \quad \text{for }i = 1,\ldots,N\,.
\end{equation}
We align a background shear flow in the same plane as the filament,
proportional in strength to the coordinate in the $\vec{e}_y$ direction,
denoted $y$. Explicitly, this flow $\vec{u}_b$ is given in the laboratory
frame in dimensionless form by
\begin{equation}
\vec{u}_b = \frac{1}{\sqrt{2}}y(\vec{e}_x + \vec{e}_z)\,,
\end{equation}
having taken the timescale $T$ to be the inverse of the dimensional shear
rate. Simulation proceeds with $E_h\approx\SI{4.7e5}{}$, consistent with the
regime found in \citet{Liu2018}, and we note that the tip of the filament
initially curves into the oncoming background flow in $y<0$.

\subsection{Relaxation of a non-planar filament}
Taking $N=100$ segments, we impose the helical initial condition
\begin{equation}
	\theta_i=\pi/3 \,, \quad \phi_i = 2\pi\frac{i-1}{N-1}\,, \quad \psi_i=0\,, \quad \text{for }i = 1,\ldots,N\,.
\end{equation}
We simulate filament motion with $E_h\approx \SI{3.1e4}{}$, with results
insensitive to this choice. The parameters used in the implementation
of \citet{Ishimoto2018a} are as in their publication, with the image system
for a plane wall accordingly removed and the actively generated torques set to
zero to allow for filament relaxation in free space. In particular, whilst the
filaments of \citet{Ishimoto2018a} are extensible, the extensional modulus of
these filaments is sufficiently high so as to provide near inextensibility in
their results, enabling meaningful comparison.

\subsection{Relaxation of a non-straight filament}
Taking $N=70$ segments, we impose the straight initial condition
\begin{equation}
	\theta_i=\pi/2 \,, \quad \phi_i = 0\,, \quad \psi_i=0\,, \quad \text{for }i = 1,\ldots,N\,.
\end{equation}
The intrinsic curvature is specified as $\vec{\kappa}^0=\pi\vec{d}_2 +
2\pi\vec{d}_3$. We simulate filament motion with $E_h\approx \SI{1.5e5}{}$,
with results insensitive to this choice.

\subsection{Active beating of a clamped filament}
Taking $N=50$ segments, we impose the straight initial condition
\begin{equation}
    \theta_i=\pi/2 \,, \quad \phi_i = 0\,, \quad \psi_i=0\,, \quad \text{for }i = 1,\ldots,N\,.
\end{equation}
The active moment density is specified as $\vec{m}^a=5\sin(s-t)\vec{d}_1 +
5\cos(s-t)\vec{d}_2$. We simulate filament motion with $E_h=10^3$ up until
$t=8\pi$.

\end{document}